\documentclass[conference]{IEEEtran}
\usepackage{cite}

\ifCLASSINFOpdf
\else
\fi

\usepackage{float}

\usepackage{graphicx}
\usepackage{epstopdf}

\usepackage{url}
\usepackage{dblfloatfix}

\DeclareGraphicsExtensions{.ps}
\begin{document}
%
\title{Towards Understanding User Preferences from User Tagging Behavior for Personalization}


\author{\IEEEauthorblockN{Amandianeze O. Nwana}
\IEEEauthorblockA{School of Electrical and Computer Engineering\\
Cornell University\\
Ithaca, NY 14850, USA \\
aon3@cornell.edu}
\and
\IEEEauthorblockN{Tsuhan Chen}
\IEEEauthorblockA{School of Electrical and Computer Engineering\\
Cornell University\\
Ithaca, NY 14850, USA \\
tsuhan@ece.cornell.edu}
}


%


\maketitle

\begin{abstract}

Personalizing image tags is a relatively new and growing area of research,
and in order to advance this research community, we must review and challenge
the de-facto standard of defining tag importance. We believe that for greater
progress to be made, we must go beyond tags that merely describe objects that
are visually represented in the image, towards more user-centric and subjective
notions such as emotion, sentiment, and preferences. 

We focus on the notion of user preferences and show that the order that users
list tags on images is correlated to the order of preference over the tags that
they provided for the image. While this observation is not completely 
surprising, to our knowledge, we are the first to explore this aspect of user
tagging behavior systematically and report empirical results to support this
observation. We argue that this observation can be exploited to help advance
the image tagging (and related) communities.

Our contributions include: 1.) conducting a user study demonstrating this
observation, 2.) collecting a dataset with user tag preferences explicitly
collected.

\end{abstract}

\begin{IEEEkeywords}
tagging; behavior; personalization; user preference
\end{IEEEkeywords}

%
\IEEEpeerreviewmaketitle

\section{Introduction}
\label{sect:intro}

\begin{figure*}[ht!]
 \includegraphics[height=.15\textheight,width=.99\textwidth]{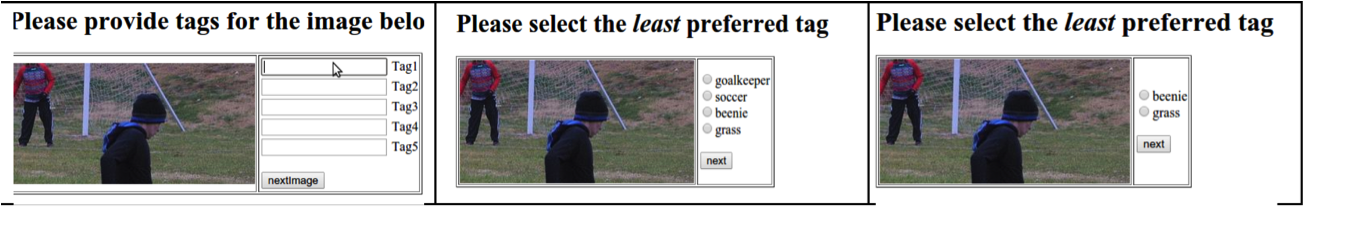}
 \caption{This figure shows screenshot of our AMT experiment. The left image,
 is from the initial tag collecting round, the middle from the second 
 tag-elimination round, and the right from the verification round.}
 \label{fig:amt_exp}
\end{figure*}

With the proliferation of cheap imaging devices (e.g., smartphone cameras,
point-and-shoots, SLRs and DSLRs) and content sharing websites (e.g., Flickr,
Tumblr, Instagram, etc), the size of personal image collections has been growing
rapidly, making it unfeasible for users to manually tag all images in their 
collections. For example, on average, 130 million images are uploaded on Tumblr
\cite{web:tumblr} and more than 90\% of those images have no identifying
text or tags\cite{web:curalate}. This makes the task of automatic image 
annotation all the more important, and with the lack of semantic understanding 
of images (``semantic gap'') this task becomes very difficult. 

Much of the work in automatic image tagging has ignored the user factor
\cite{gong:13,li:08, li:11,lin:13,rubenstein:12,tomasik:09,wang:11,weston:11,
wray:10, zhu:10} by trying to find what we denote as statistical correlations 
between the image content (visual features) and objective semantics regardless
of the particular users involved in the tagging activity. 
There has been some work that focuses on user personalization in automated
image tagging, most notably, \cite{li:11}, which we consider as the
state-of-art in this domain. Along the lines of object importances as they relate
to tags, the focus has mainly been on an explicitly categorical definition of
importance by measuring properties of content/objects in the image (e.g., size, 
salience, etc) to estimate their relative importances \cite{berg:12}. These
content property based approaches to importance also tends to ignore particular
user effects and preferences, treating importance as a purely global phenomenon
\cite{berg:12, spain:11}. In our day and age where content is increasingly
personalized and tailored to user tastes, we believe that it is of paramount
importance to systematically understand user tagging behavior and trends.

A recent attempt at a design change on Flickr~\cite{web:flickr}, and the 
subsequent reversal of the change, demonstrates our second assumption. The Flickr
designers opted to update the site to present user generated tags in 
reverse-chronological order, and immediately active Flickr users protested
this change, citing that the order that they presented their tags was
intentional~\cite{web:flickr_order}, leading to an apology by the designers
and a reverting back to the original chronological order design. This event
lead us to believe when providing tag lists, users are not merely motivated
by visually measurable properties such as saliency, but more so by implicit
biases and preferences which are in turn reflected in the order of tagging list.



In the subsequent sections, we investigate our aforementioned hypotheses via a
user study conducted using the Amazon Mechanical Turk (AMT) system, and compare
to more popular global notions of importance.


\subsection{Related Works}
\label{sec:related}

There are two ways to approach image tagging. First, explicit
object tagging, where an image is tagged with a particular word if the object
the word represents is detected as being in the image. Second, implicit tagging, where the query
image is compared to other similar images, and the tags are ``transferred'' 
from the most similar images to the query image, via some scoring function.
Many applications of this implicit approach take their cue from the world of 
collaborative filtering \cite{weston:11, li:10}. 

The implicit approaches are usually more common than their explicit counterparts
because one does not have to learn how to recognize or detect specific objects
in the image, which as earlier noted is not scalable, also not all concepts one 
would like to use in describing an image are necessarily visual (semantic gap)
\cite{smeulders:00, schreiber:01}.  Also,
with the implicit approaches one could imagine a latent space that more readily
embeds some sense of relatedness \cite{weston:11}, while on the explicit end, it is harder to 
extrapolate a measure of relatedness between objects of different classes. 

With regards to personalization in image tagging, in the work by Rendle et al.
\cite{rendle:10}, they assume that tag mentioned are preferred to those unmentioned. This is
similar to our assumption but they treat the tags that appear \emph{together} on a tagging
list equally, which our work here suggests not to do. In the work by Lipczak et al.~\cite{lipczak:11} they also treat
tags as essentially structureless entities (bag of words), and other work on personalization
similarly treat user provided tag lists as such~\cite{lipczak:11, li:11}. To our knowledge,
we are the first to suggest that these user provided list should be treated as having
structure. 


\section{User Study}
\label{sec:user_study}

In order to verify our assumption that users tend to place the tags they
prefer or consider most important at the start of tag lists, we found it
prudent to conduct a live user experiment using the AMT system. Our main
metric of interest is the rank correlation between tags lists when we
explicitly request and ascertain the preference order of tags they provided
for an image, versus the order of the same tags without such a prompt for
ranking their preferences. In the following, we will detail the setup of our
user study, our metrics and measurements, comparison to other measures of 
(global) importance, and our conclusions from the user study.

\subsection{Study Setup}
\label{sec:study_setup}

We conducted our study on a subset of 500 images from the NUS-WIDE dataset~\cite{
nus-wide-civr09} which is a dataset of images from the popular photoblogging 
website Flickr~\cite{web:flickr}. These images were divided into 100 groups of 
5 to create 100 Human Intelligence Tasks (HITs), which is the smallest 
indivisible unit of work on AMT. Each HIT was then assigned to 15 different 
study participants (turkers), totaling 1500 assignments. 

For each HIT, our study was done in 2 stages as shown in \figurename~\ref{fig:amt_exp}. 
In the first stage, we asked the turker to provide 5 tags for each of the 5
images contained in the HIT, we refer to this as the {\bf\textit{Tag Allocation Stage}}. 
The tags are allocated for all of the 5 images before we begin the next stage of the
experiment.
In the second stage of the study, we iteratively asked the turker to eliminate their
\emph{least} preferred tag for each image from the set of tags that had not yet been
eliminated for that image in previous iterations. For each round of elimination, 
we randomly scrambled the order of the \emph{tags} that were left from previous
rounds to prevent turkers from any influence of presentation bias. Similarly, the
order of the \emph{images} in each elimination round was randomly shuffled to 
prevent presentation bias. We refer to the second stage as the 
{\bf\textit{Preference Allocation Stage}}.

We also added a hidden verification test as part of the preference allocation stage to 
ensure that the preferences which the turker provided were consistent when
asked a second time. To that end, after reconstructing their preference order
from 4 elimination rounds, for each image, we asked the turker to eliminate
their least preferred tag among 2 randomly chosen tags of those which the user
had provided for that image. \emph{If their response matched their reconstructed 
preference order, then we considered the user's preference order for that image
verified otherwise not}. So within each HIT we can tell which of the 
reconstructed preference lists are reliable, and so on the level of HITs we can
define the HIT reliability as the number of verified preference orders within
the HIT.

At the end of the experiment we are left with 7500 pairs of tag lists from 391
turkers. Each tag list pair consists of a tag list in the default order and the
same tags in the user's preferred order. And for each of those pairs we know
whether or not the preferred order is verified and use that as a proxy to its
reliability.

\subsection{Metrics and Measurements}
\label{sec:study_metrics}

\begin{figure*}[th!]
 \includegraphics[height=.3\textheight,width=.99\textwidth]{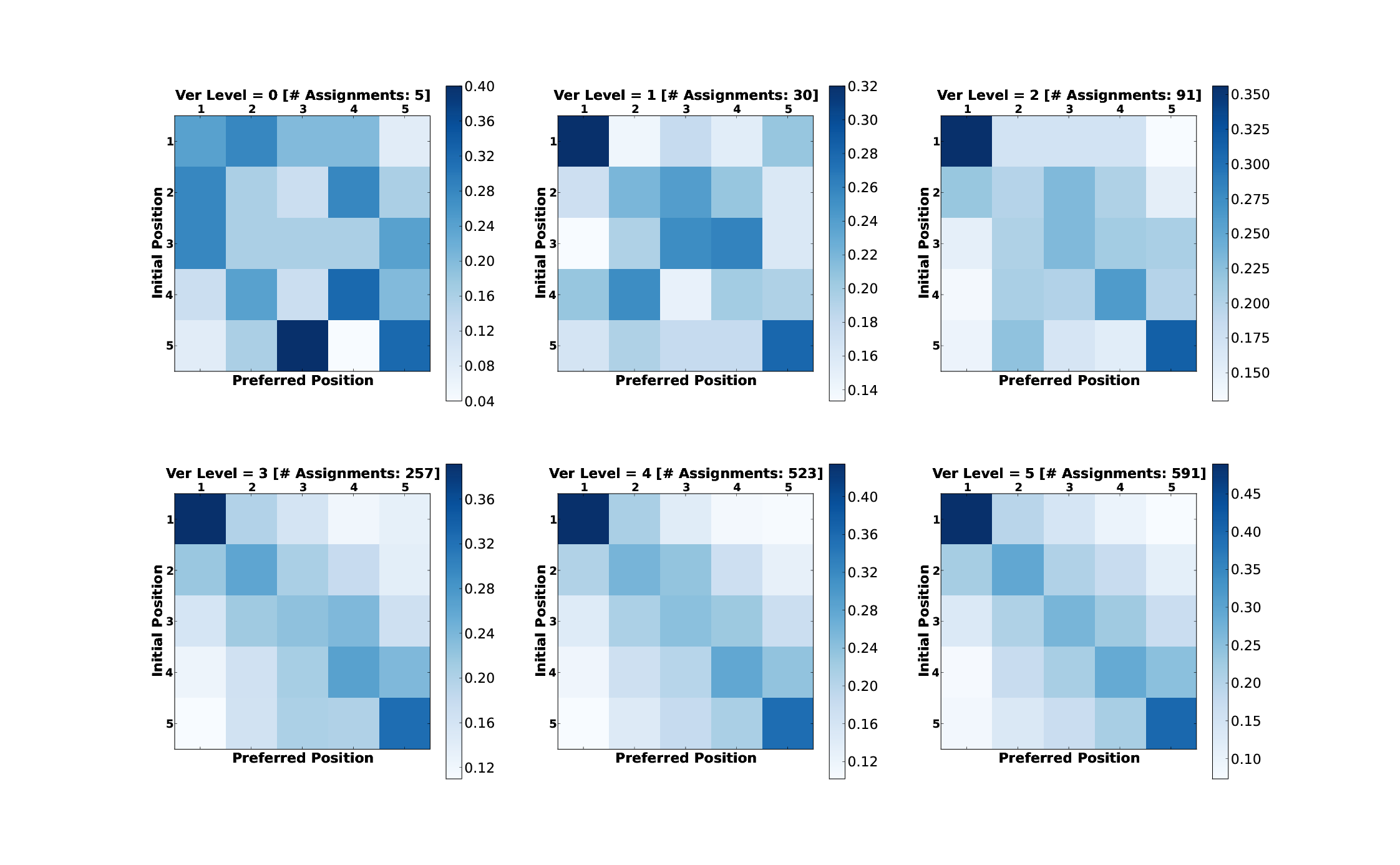}
 \caption{This figure shows the confusion matrix between the preferred position of tags
 (groundtruth label) and the initial position of tags (predicted label) at 
 different levels of verification per HIT. The verification level is the number of
 images (out of 5) that were verified within the hit.}
 \label{fig:conf_mat}
\end{figure*}

\begin{table*}[th]
  \caption{}
    \label{tab:corr_stats}
  \begin{center}
    \begin{tabular}{| l || c | c | c | c | c | c |}
    \hline
              & Avg. $\tau$ corr & Var. $\tau$ corr  & Avg. $\rho$~corr & Var. $\rho$ corr & min. verification/total & num verified/total \\
    \hline\hline
    Per Assignment   & 0.3089   & 0.062  & 0.3705  & 0.083 & 4/5 & 1114/1500  \\
    \hline
    Per User  & 0.3046 & 0.047  & 0.3652   & 0.064 & 3.5/5 & 298/391 \\
    \hline
    Per Image & 0.2840 & 0.017  & 0.3400   & 0.021 & 11/15 & 434/500 \\
    \hline
    \end{tabular}
  \end{center}
  This table shows the correlation statistics between the initial rank of tags,
  and the preference ranks as provided by the user. We provide the averages per assignment,
  per user (averaging over all images for the user), and per image (averaging over all users
  for that image).
\end{table*}

To verify our assumption that users tend to present their tag lists with an
inherent preference order as opposed to being an orderless set or bag-of-words,
we examine the data collected from our AMT user study. To measure whether the
data supports our claim, we employ 2 metrics: 1.) Kendall's Tau Rank 
Correlation~\cite{kendall_tau}, and 2.) Spearman's Rho Rank Correlation~\cite{
spearman_rho}, which are both measures of how much two rankings are correlated
with one another. Both measures range from -1 to 1, with -1 indicating perfect
negative correlation, 0 indicating no correlation, and 1 indicating perfect
positive correlation. 

We measure the average correlation per user, and the average correlation per
image. We also measure the effect of the verification of the preference order
on the correlation scores, and report our final numbers based on data that has
been reasonably verified. We also present the confusion matrix between the 
position ``labels'', that is, assuming each tag is labeled with it's position
from the initial order, how well does its position ``label'' on the preference 
order list predict its position ``label'' from the initial order.

As we can see from the confusion matrices in \figurename~\ref{fig:conf_mat}, assuming
that the verification level (here as the number of images verified within a
single HIT) \emph{is} a proxy of the turker's attention to the task (and hence 
reliability), the position of a tag on the initial list is a good predictor
of the position of the tag according to the turker's preference. As the 
reliability increases, more often than not, the initial position is the same as
the preferred position, and any ``mislabeling'' is typically within an error of
1 position. 

\begin{table*}[th]
 \caption{}
 \label{tab:freq_stats}
  \begin{center}
    \begin{tabular}{| l || c | c | c | c | c |}
    \hline
              & Avg. $\tau$ corr & Var. $\tau$ corr  & Avg. $\rho$~corr & Var. $\rho$ corr\\
    \hline\hline
    Overall    & 0.187417923625   & 0.177135465861  & 0.220259416265  & 0.232534427934   \\
    \hline
    {Image (avg. over users)} & 0.186084678459 & 0.0230843379766  & 0.218679398053   & 0.0306837413827  \\
    \hline
    \end{tabular}
  \end{center}
  This table shows the correlation statistics between the frequency rank of tags,
  and the preference ranks as provided by the user. The frequency rank of a tag for an image is
  derived from the number of times it was mentioned by all the turkers that tagged the given
  image. We provide the correlation statistics over all the tag list, and also averaged across 
  the users for each image. We only report the statistics for images that were verified, using
  all the images results in even lower correlation.
\end{table*}

\begin{figure*}[ht!]
 \includegraphics[height=.28\textheight,width=.99\textwidth]{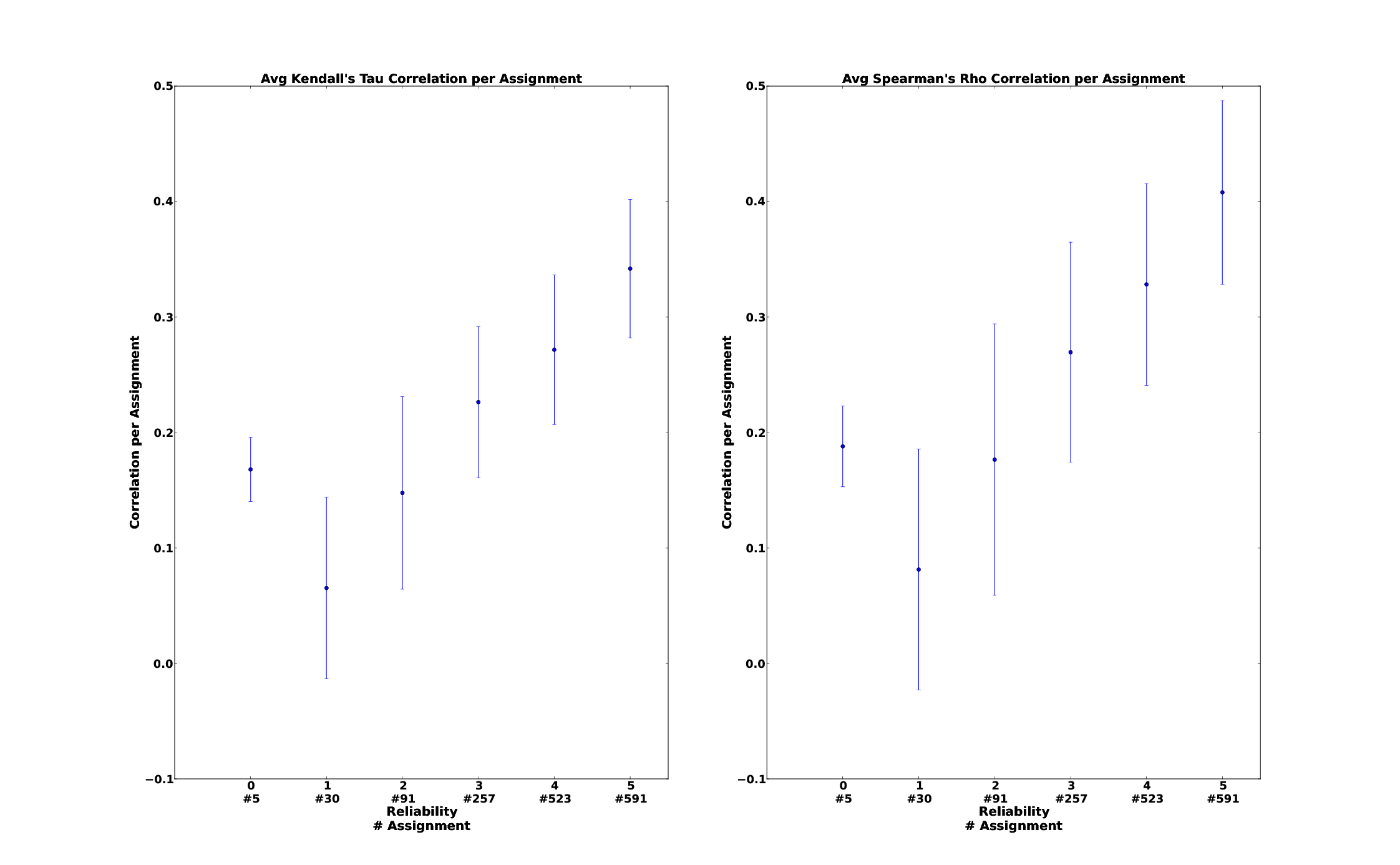}
 \caption{This figure shows the average correlation (and error bars) between the
 initial tag list, and the reconstructed preference order with respect to the 
 level of reliability. The Kendall's Tau correlation is shown on the left, and
 Spearman's Rho on the right.}
 \label{fig:corr_asgn}
\end{figure*}

In \figurename~\ref{fig:corr_asgn} we see that from the reliable HITs, there seems
to be a moderately high correlation (which is \emph{statistically significantly}
different from being uncorrelated, as validated by a  two-sided 1 sample t-test
with p-values less than 0.01) between the initial tag list, and the 
reconstructed preference order, and we believe we are the first to show 
empirically, the existence of such phenomenon, which apriori is not so obvious. 

\begin{figure*}[th!]
 \includegraphics[height=.3\textheight,width=.99\textwidth]{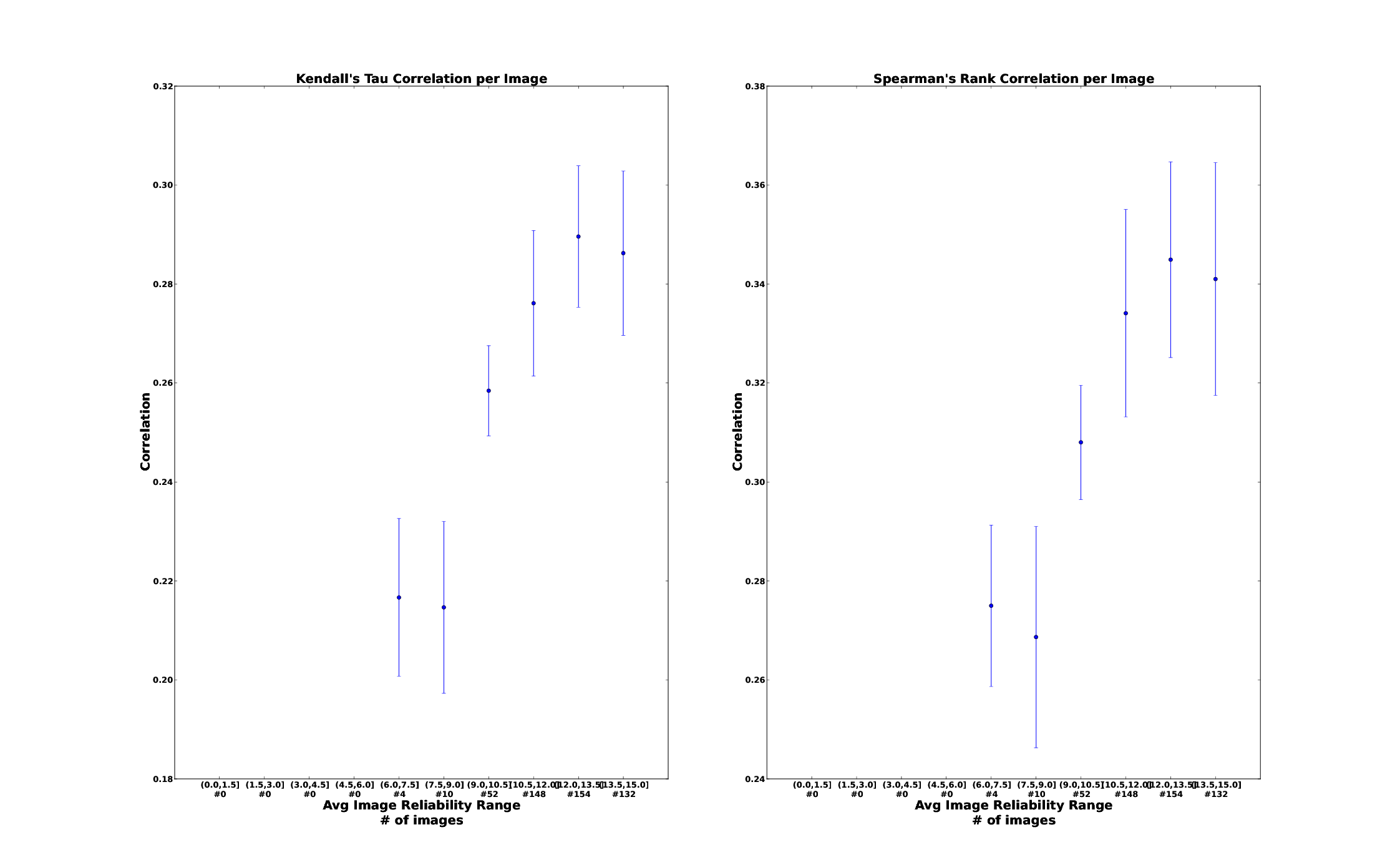}
 \caption{This figure shows the average correlation per image (and error bars) 
 between the initial tag list, and the reconstructed preference order with 
 respect to the number of times the image has been reliably tagged. 
 The Kendall's Tau correlation is shown on the left, and Spearman's Rho on the
 right.}
 \label{fig:corr_img}
\end{figure*}

\begin{figure*}[ht!]
 \includegraphics[height=.3\textheight,width=.99\textwidth]{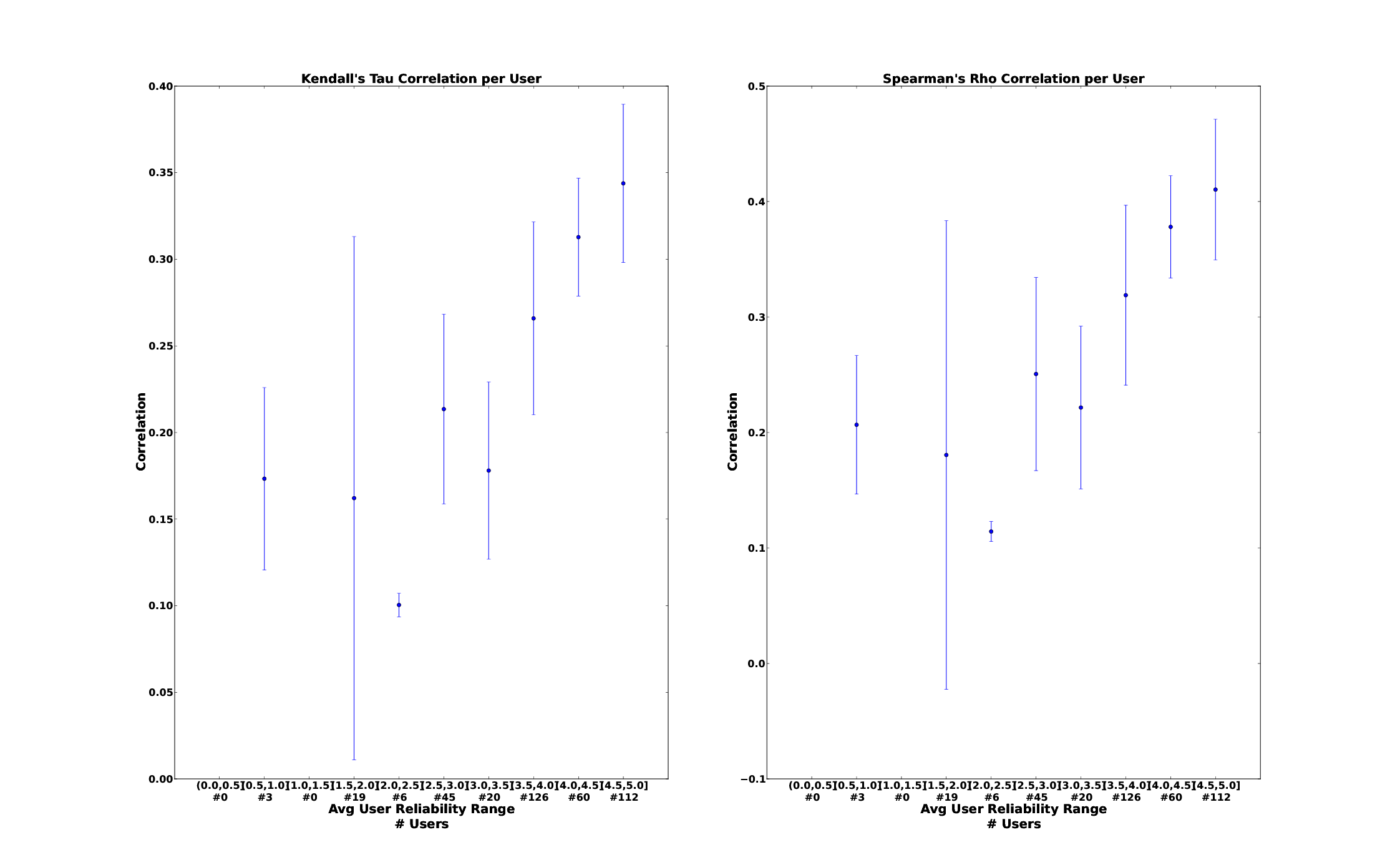}
 \caption{This figure shows the average correlation (and error bars) between the
 initial tag list, and the reconstructed preference order with respect to the 
 level of reliability. The Kendall's Tau correlation is shown on the left, and
 Spearman's Rho on the right.}
 \label{fig:corr_user}
\end{figure*}

Each image is tagged by 15 different turkers, and most of the images, more often
than not, were tagged reliably by the turkers. From \figurename~\ref{fig:corr_img} we
can see that the aforementioned correlation is largely independent of the image,
as even those images with less reliable preference orders show a moderate 
correlation between the reconstructed preference tag order and the initial tag
list, so it doesn't seem to be the case that image visual content itself is the cause of the
correlation. 
When we consider the average correlation per user as is shown in
\figurename~\ref{fig:corr_user} we also observe the similar trend that users that have
tagged images more reliably show on average a moderately high correlation
between the reconstructed preference order, and even the less reliable users
still exhibit a slight correlation as well. Our results are summarized in
Table~\ref{tab:corr_stats}, and these are statistically significant as 
verified by a one sample t-test with respect to 0 correlation.

\subsection{Comparison to Global Importance}
\label{sec:study_comparsion}

In much of image tagging research~\cite{li:11, hwang:12, spain:11, liu:09,berg:12}, tag
importance is usually considered in terms of what is visually represented in the
image, and typically by saliency. To that end, many researchers use tag 
frequency as a proxy to tag importance and saliency~\cite{hwang:12, spain:11}, and for nearest neighbor
approaches to tagging, predicting the tags that are based on the most frequent
has had relative success in terms of tag recall~\cite{li:11}.

In this section, we compare the reconstructed user preference order to the 
frequency order gotten from the number of times the tag was mentioned by
turkers for that image. In Table~\ref{tab:freq_stats}, we report the correlation
statistics. As we can see, there is a slight correlation between the preference
and the frequency, but it is not that strong, which suggests that although users
might mention tags of global importance (or salient tags) in their tag lists, 
those tags are usually not their most preferred. 

In order to verify that
suggestion, we also report the average position of the most frequently mentioned
tag for an image on the reconstructed preference ordered list for the same 
image and notice that more often than not, the most frequent tag is usually
mentioned later in the preference order as is seen in Table~\ref{tab:pos_stats}.

\subsection{Study Summary}
\label{sec:study_conclusions}

From our study we arrive at the following conclusions: 1.) The order that
users provide in their tag list for an image is moderately correlated to their
inherent preferences over those tags, 2.) This preferred order is not as simple
as the order of objects in the image from most salient to least salient, nor
the same as other global notions of preference, and 3.) Hence in understanding
user tagging behavior and inferring user preferences, one should consider the
order that users present their tags for images. 

We believe that this study will help further the development of research in the
area of image tagging, and that using the observations provided by this study,
could improve upon current state-of-art methods for image tagging, 
especially with respect to personalization. 

\begin{table}[h!]
  \caption{}
  \label{tab:pos_stats}
  \begin{center}
    \begin{tabular}{| l || c | c | c | c | c |}
    \hline
				      & Average 	& Variance  \\
    \hline\hline
    Position of most frequent tag    & 3.7652         & 0.356735173152 \\
    \hline
    \end{tabular}
  \end{center}
  This table shows the average position of the most frequently mentioned
  tag for an image, and its variance, in the preference list given by the users.
  As we can see out of the 5 tags given by the user, the most frequently mentioned
  ones tend to be closer to the bottom of the list, i.e., less preferred. We only report
  the statistics for images that were verified.
\end{table}

\section{Conclusion and Future Work}

In this work we proposed a new measurement of tag preferences, and
demonstrated that there is indeed a tag-order bias, that is, when a user
mentions tag $a$ before tag $b$, in a list of tags for a given image, the
user is implying that he/she prefers, or considers $a$ to be of greater importance
than $b$. This leads us to conclude that although there are many visual factors
that may affect what tags a user will provide for an image, it is more useful
to characterize instead (or rather in conjunction with) the users' tagging
habits to learn what tags are of more importance to the users, whether
visually motivated or not, and automatic tagging systems should
employ this technique to improve their overall performance. 

It is also important to note that this study was not tied to any particular 
online tagging system, like Flickr, and as such we believe that the findings
in this study are independent of the online platform, as opposed to being an
artifact of the user interface. Hence, the findings should hold on most online
tagging systems, or at least image tagging systems that allow for user input via text. 
One direct way we believe this preference information can be exploited is,
given a user's tagging history, if tags $a$ and $b$ frequently occur on the 
same tag lists for images, and tag $a$ is mentioned before $b$ more often than
the reverse, in predicting a tag list for a new image for that user, this preference
order should be enforced as it reflects a preference for $a$ over $b$ for that user.

Another future direction, assuming we can embed the tags into some metric space, is, we
believe it would be interesting to learn a function that takes as input, a pair of
features (each representing a tag) and returns a prediction of the pair
preference order and strength. This will enable us to ``transfer'' preferences
between tags that are similar (or closely related) even though we might never
have observed them together for a particular user. We would also like to 
analyze what kinds/categories of tags are preferred over others under this framework,
and answer the question, do these categorical relationships depend on the user 
(i.e., do the users cluster in a way such that the different clusters exhibit 
different categorical relationships)? For example do some user tend to tag
images in a bottom-up fashion with respect to ontologies, and other users in
a top-bottom fashion?


%
%



\bibliographystyle{IEEEtran}
\bibliography{ism15}

\begin{thebibliography}{10}
\providecommand{\url}[1]{#1}
\csname url@samestyle\endcsname
\providecommand{\newblock}{\relax}
\providecommand{\bibinfo}[2]{#2}
\providecommand{\BIBentrySTDinterwordspacing}{\spaceskip=0pt\relax}
\providecommand{\BIBentryALTinterwordstretchfactor}{4}
\providecommand{\BIBentryALTinterwordspacing}{\spaceskip=\fontdimen2\font plus
\BIBentryALTinterwordstretchfactor\fontdimen3\font minus
  \fontdimen4\font\relax}
\providecommand{\BIBforeignlanguage}[2]{{%
\expandafter\ifx\csname l@#1\endcsname\relax
\typeout{** WARNING: IEEEtran.bst: No hyphenation pattern has been}%
\typeout{** loaded for the language `#1'. Using the pattern for}%
\typeout{** the default language instead.}%
\else
\language=\csname l@#1\endcsname
\fi
#2}}
\providecommand{\BIBdecl}{\relax}
\BIBdecl

\bibitem{web:tumblr}
``Tumblr statistics,''
  \url{http://socialtimes.com/tumblr-scan-images-brand-content_b202520}, last
  Accessed: 10/26/2014.

\bibitem{web:curalate}
``Curalte statistics,'' \url{http://www.wired.com/2014/06/curalate/}, last
  Accessed: 10/26/2014.

\bibitem{gong:13}
Y.~Gong, Y.~Jia, T.~Leung, A.~Toshev, and S.~Ioffe, ``Deep convolutional
  ranking for multilabel image annotation,'' \emph{arXiv preprint
  arXiv:1312.4894}, 2013.

\bibitem{li:08}
\BIBentryALTinterwordspacing
X.~Li, C.~G. Snoek, and M.~Worring, ``Learning tag relevance by neighbor voting
  for social image retrieval,'' in \emph{Proceedings of the 1st ACM
  International Conference on Multimedia Information Retrieval}, ser. MIR
  '08.\hskip 1em plus 0.5em minus 0.4em\relax New York, NY, USA: ACM, 2008, pp.
  180--187. [Online]. Available:
  \url{http://doi.acm.org/10.1145/1460096.1460126}
\BIBentrySTDinterwordspacing

\bibitem{li:11}
\BIBentryALTinterwordspacing
X.~Li, E.~Gavves, C.~G. Snoek, M.~Worring, and A.~W. Smeulders, ``Personalizing
  automated image annotation using cross-entropy,'' in \emph{Proceedings of the
  19th ACM International Conference on Multimedia}, ser. MM '11.\hskip 1em plus
  0.5em minus 0.4em\relax New York, NY, USA: ACM, 2011, pp. 233--242. [Online].
  Available: \url{http://doi.acm.org/10.1145/2072298.2072330}
\BIBentrySTDinterwordspacing

\bibitem{lin:13}
Z.~Lin, G.~Ding, M.~Hu, J.~Wang, and X.~Ye, ``Image tag completion via
  image-specific and tag-specific linear sparse reconstructions,'' in
  \emph{Computer Vision and Pattern Recognition (CVPR), 2013 IEEE Conference
  on}, June 2013, pp. 1618--1625.

\bibitem{rubenstein:12}
\BIBentryALTinterwordspacing
M.~Rubinstein, C.~Liu, and W.~T. Freeman, ``Annotation propagation in large
  image databases via dense image correspondence,'' in \emph{Proceedings of the
  12th European Conference on Computer Vision - Volume Part III}, ser.
  ECCV'12.\hskip 1em plus 0.5em minus 0.4em\relax Berlin, Heidelberg:
  Springer-Verlag, 2012, pp. 85--99. [Online]. Available:
  \url{http://dx.doi.org/10.1007/978-3-642-33712-3_7}
\BIBentrySTDinterwordspacing

\bibitem{tomasik:09}
\BIBentryALTinterwordspacing
B.~Tomasik, P.~Thiha, and D.~Turnbull, ``Tagging products using image
  classification,'' in \emph{Proceedings of the 32Nd International ACM SIGIR
  Conference on Research and Development in Information Retrieval}, ser. SIGIR
  '09.\hskip 1em plus 0.5em minus 0.4em\relax New York, NY, USA: ACM, 2009, pp.
  792--793. [Online]. Available:
  \url{http://doi.acm.org/10.1145/1571941.1572131}
\BIBentrySTDinterwordspacing

\bibitem{wang:11}
Y.~Wang and G.~Mori, ``Max-margin latent dirichlet allocation for image
  classification and annotation,'' in \emph{Proceedings of the British Machine
  Vision Conference}.\hskip 1em plus 0.5em minus 0.4em\relax BMVA Press, 2011,
  pp. 112.1--112.11, http://dx.doi.org/10.5244/C.25.112.

\bibitem{weston:11}
\BIBentryALTinterwordspacing
J.~Weston, S.~Bengio, and N.~Usunier, ``Wsabie: Scaling up to large vocabulary
  image annotation,'' in \emph{Proceedings of the Twenty-Second International
  Joint Conference on Artificial Intelligence - Volume Volume Three}, ser.
  IJCAI'11.\hskip 1em plus 0.5em minus 0.4em\relax AAAI Press, 2011, pp.
  2764--2770. [Online]. Available:
  \url{http://dx.doi.org/10.5591/978-1-57735-516-8/IJCAI11-460}
\BIBentrySTDinterwordspacing

\bibitem{wray:10}
T.~Wray and P.~W. Eklund, ``Social tagging for digital libraries using formal
  concept analysis,'' 2010.

\bibitem{zhu:10}
\BIBentryALTinterwordspacing
G.~Zhu, S.~Yan, and Y.~Ma, ``Image tag refinement towards low-rank, content-tag
  prior and error sparsity,'' in \emph{Proceedings of the International
  Conference on Multimedia}, ser. MM '10.\hskip 1em plus 0.5em minus
  0.4em\relax New York, NY, USA: ACM, 2010, pp. 461--470. [Online]. Available:
  \url{http://doi.acm.org/10.1145/1873951.1874028}
\BIBentrySTDinterwordspacing

\bibitem{berg:12}
A.~Berg, T.~Berg, H.~Daume, J.~Dodge, A.~Goyal, X.~Han, A.~Mensch, M.~Mitchell,
  A.~Sood, K.~Stratos, and K.~Yamaguchi, ``Understanding and predicting
  importance in images,'' in \emph{Computer Vision and Pattern Recognition
  (CVPR), 2012 IEEE Conference on}, June 2012, pp. 3562--3569.

\bibitem{spain:11}
\BIBentryALTinterwordspacing
M.~Spain and P.~Perona, ``Measuring and predicting object importance,''
  \emph{Int. J. Comput. Vision}, vol.~91, no.~1, pp. 59--76, Jan. 2011.
  [Online]. Available: \url{http://dx.doi.org/10.1007/s11263-010-0376-0}
\BIBentrySTDinterwordspacing

\bibitem{web:flickr}
``Flickr,'' \url{http://www.flickr.com}, last Accessed: 01/26/2015.

\bibitem{web:flickr_order}
``On flickr ordering,''
  \url{https://www.flickr.com/help/forum/en-us/72157645219834187/}, last
  Accessed: 01/26/2015.

\bibitem{li:10}
\BIBentryALTinterwordspacing
X.~Li, C.~G.~M. Snoek, and M.~Worring, ``Unsupervised multi-feature tag
  relevance learning for social image retrieval,'' in \emph{Proceedings of the
  ACM International Conference on Image and Video Retrieval}, ser. CIVR
  '10.\hskip 1em plus 0.5em minus 0.4em\relax New York, NY, USA: ACM, 2010, pp.
  10--17. [Online]. Available: \url{http://doi.acm.org/10.1145/1816041.1816044}
\BIBentrySTDinterwordspacing

\bibitem{smeulders:00}
A.~W.~M. Smeulders, M.~Worring, S.~Santini, A.~Gupta, and R.~Jain,
  ``Content-based image retrieval at the end of the early years,'' \emph{IEEE
  TRANSACTIONS ON PATTERN ANALYSIS AND MACHINE INTELLIGENCE}, vol.~22, no.~12,
  pp. 1349--1380, 2000.

\bibitem{schreiber:01}
A.~T.~G. Schreiber, B.~Dubbeldam, J.~Wielemaker, and B.~Wielinga,
  ``Ontology-based photo annotation,'' 2001.

\bibitem{rendle:10}
\BIBentryALTinterwordspacing
S.~Rendle and L.~Schmidt-Thieme, ``Pairwise interaction tensor factorization
  for personalized tag recommendation,'' in \emph{Proceedings of the Third ACM
  International Conference on Web Search and Data Mining}, ser. WSDM '10.\hskip
  1em plus 0.5em minus 0.4em\relax New York, NY, USA: ACM, 2010, pp. 81--90.
  [Online]. Available: \url{http://doi.acm.org/10.1145/1718487.1718498}
\BIBentrySTDinterwordspacing

\bibitem{lipczak:11}
\BIBentryALTinterwordspacing
M.~Lipczak and E.~Milios, ``Efficient tag recommendation for real-life data,''
  \emph{ACM Trans. Intell. Syst. Technol.}, vol.~3, no.~1, pp. 2:1--2:21, Oct.
  2011. [Online]. Available: \url{http://doi.acm.org/10.1145/2036264.2036266}
\BIBentrySTDinterwordspacing

\bibitem{nus-wide-civr09}
T.-S. Chua, J.~Tang, R.~Hong, H.~Li, Z.~Luo, and Y.-T. Zheng, ``Nus-wide: A
  real-world web image database from national university of singapore,'' in
  \emph{Proc. of ACM Conf. on Image and Video Retrieval (CIVR'09)}, Santorini,
  Greece., July 8-10, 2009.

\bibitem{kendall_tau}
\BIBentryALTinterwordspacing
M.~G. Kendall, ``\BIBforeignlanguage{English}{A new measure of rank
  correlation},'' \emph{\BIBforeignlanguage{English}{Biometrika}}, vol.~30, no.
  1/2, pp. pp. 81--93, 1938. [Online]. Available:
  \url{http://www.jstor.org/stable/2332226}
\BIBentrySTDinterwordspacing

\bibitem{spearman_rho}
\BIBentryALTinterwordspacing
C.~Spearman, ``\BIBforeignlanguage{English}{The proof and measurement of
  association between two things},'' \emph{\BIBforeignlanguage{English}{The
  American Journal of Psychology}}, vol.~15, no.~1, pp. pp. 72--101, 1904.
  [Online]. Available: \url{http://www.jstor.org/stable/1412159}
\BIBentrySTDinterwordspacing

\bibitem{hwang:12}
\BIBentryALTinterwordspacing
S.~J. Hwang and K.~Grauman, ``Learning the relative importance of objects from
  tagged images for retrieval and cross-modal search,'' \emph{Int. J. Comput.
  Vision}, vol. 100, no.~2, pp. 134--153, Nov. 2012. [Online]. Available:
  \url{http://dx.doi.org/10.1007/s11263-011-0494-3}
\BIBentrySTDinterwordspacing

\bibitem{liu:09}
\BIBentryALTinterwordspacing
D.~Liu, X.-S. Hua, L.~Yang, M.~Wang, and H.-J. Zhang, ``Tag ranking,'' in
  \emph{Proceedings of the 18th International Conference on World Wide Web},
  ser. WWW '09.\hskip 1em plus 0.5em minus 0.4em\relax New York, NY, USA: ACM,
  2009, pp. 351--360. [Online]. Available:
  \url{http://doi.acm.org/10.1145/1526709.1526757}
\BIBentrySTDinterwordspacing

\end{thebibliography}
%

\end{document}